\documentclass[prd,twocolumn,nopacs,floatfix,amsmath,nofootinbib,amssymb]{revtex4}
\usepackage{graphicx,color,dcolumn,booktabs,bm}
\usepackage{longtable,lscape}
\usepackage{txfonts}
\usepackage{overpic}
\usepackage{amssymb}
\usepackage{diagbox}
\usepackage{multirow}
\usepackage{booktabs}
\usepackage{indentfirst}
\usepackage{feynmf}   
\usepackage{slashed}  
\usepackage{cases}
\usepackage{color}
\usepackage{multirow}
\usepackage{epstopdf}
\usepackage{graphicx,color,dcolumn,booktabs,bm}
\usepackage[colorlinks, citecolor=blue,anchorcolor=red,menucolor=red, linkcolor=red,filecolor=red,runcolor=red,urlcolor=blue,frenchlinks=red]{hyperref}

\begin{document}

\title{Interpretation of the excited $\Omega_c$ family via mass and width in the chiral quark model}

\author{Xiao-Huang Hu$^1$}\email{huxiaohuang@ciit.edu.cn}
\author{Zhe-Tao Miu$^2$}
\author{Yue Tan$^3$}\email{tanyue@ycit.edu.cn}
\author{Qi Huang$^{2,5}$}\email{06289@njnu.edu.cn}
\author{Ye Yan$^4$}
\author{Jia-Lun Ping$^2$}
\affiliation{
$^1$Department of Physics,Changzhou Institute of Industry Technology, Changzhou 213164, China\\
$^2$Department of Physics and Technology, Nanjing Normal University, Nanjing 210023, China\\
$^3$Department of Physics, Yancheng Institute of Technology, Yancheng 224000, China\\
$^4$Department of Physics, Changzhou University of Information Technology, Changzhou 213164, China\\
$^5$Lanzhou Center for Theoretical Physics, Key Laboratory of Theoretical Physics of Gansu Province, Lanzhou University, Lanzhou 730000, China}

\begin{abstract}
In this work, the mass and decay properties for low-lying orbital excitations of $\Omega_c$ baryons are investigated within the chiral quark model combined with the $^3P_0$ decay mechanism. Following the LHCb Collaboration's observations, we extend our previous study of the $\Xi_c^{\prime}$ system to the $\Omega_c$ system, with all model parameters inherited from our earlier work. Our results consistently describe the observed $\Omega_c$ spectrum:
$\Omega_c(3000)$, $\Omega_c(3050)$, $\Omega_c(3065)$, and $\Omega_c(3090)$ 
are well described as $\lambda$-mode $1P$ states with $J^P=1/2^-$, $3/2^-$, $3/2^-$, and $5/2^-$, respectively. $\Omega_c(3119)$ cannot be described as a pure three-quark excitation. Furthermore, the $\Omega_c(3185)$ and $\Omega_c(3327)$ may be assigned to the $2S$ and $1D$ states, respectively. These results offer a coherent picture of the low-lying $\Omega_c$ spectrum and motivate further experimental tests.

\end{abstract}

\maketitle

\setcounter{totalnumber}{5}

\section{\label{sec:introduction}Introduction}

As the transition area of quantum chromodynamics (QCD), hadrons with charm quarks play important roles in decoding the non-perturbative property of QCD. Among these states, singly charmed baryon is an interesting kind of system. Compared with charmed mesons, singly charmed baryons consist of one heavy quark and two correlated light quarks. Thus, they are not only the simplest multiquark systems, but also their spectra are sensitive to the interactions between the heavy quark and the light degrees of freedom, in addition with the correlations between the two light quarks~\cite{Cheng:2021qpd,Chen:2022asf}. Among these singly charmed baryons, the $\Omega_c$ baryons, composed of one charm quark and two strange quarks, constitute a particularly distinctive system. Since the two light quarks are identical strange quarks, their flavor wave function is necessarily symmetric, while the Pauli principle imposes additional constraints on the allowed spin-space configurations. Unlike the $\Xi_c^{(\prime)}$ family, which contains both flavor-antitriplet and flavor-sextet light-diquark configurations, the $\Omega_c$ baryons have a simpler flavor structure and fewer allowed configurations. Consequently, the $\Omega_c$ spectrum provides a comparatively clean platform for investigating the dynamics between the heavy quark and the light degrees of freedom and for testing QCD-inspired quark models.

The ground-state $\Omega_c(2695)^0$ with $J^P=1/2^+$ was first established in 1985~\cite{Biagi:1984mu}, while its spin partner $\Omega_c(2770)^0$ with $J^P=3/2^+$ was first observed by the BABAR Collaboration in 2006~\cite{BaBar:2006pve} and subsequently studied by the Belle Collaboration~\cite{Solovieva:2008fw}. Apart from these two low-lying states, no additional excited $\Omega_c$ baryons were observed for many years. A major experimental breakthrough was achieved in 2017, when the LHCb Collaboration observed five narrow excited states, $\Omega_c(3000)^0$, $\Omega_c(3050)^0$, $\Omega_c(3065)^0$, $\Omega_c(3090)^0$, and $\Omega_c(3119)^0$, in the $\Xi_c^+K^-$ invariant-mass spectrum, with the central values of their measured decay widths all below $10~\mathrm{MeV}$~\cite{LHCb:2017uwr}. The Belle Collaboration subsequently confirmed the first four states, whereas no significant signal was found for $\Omega_c(3119)^0$~\cite{Belle:2017ext}. In 2021, the four Belle-confirmed states were observed by LHCb in the decay $\Omega_b^-\to\Xi_c^+K^-\pi^-$~\cite{LHCb:2021ptx}. The hypotheses $J=1/2$ for $\Omega_c(3050)^0$ and $\Omega_c(3065)^0$ were disfavored with significances of $2.2\sigma$ and $3.6\sigma$, respectively. More recently, LHCb confirmed all five structures and observed two additional states, $\Omega_c(3185)^0$ and $\Omega_c(3327)^0$, in the same $\Xi_c^+K^-$ final state~\cite{LHCb:2023sxp}. Nevertheless, the spin-parity quantum numbers of these states have not yet been experimentally determined.

\begin{figure}[ht]
\begin{center}
\includegraphics[width=0.46\textwidth]{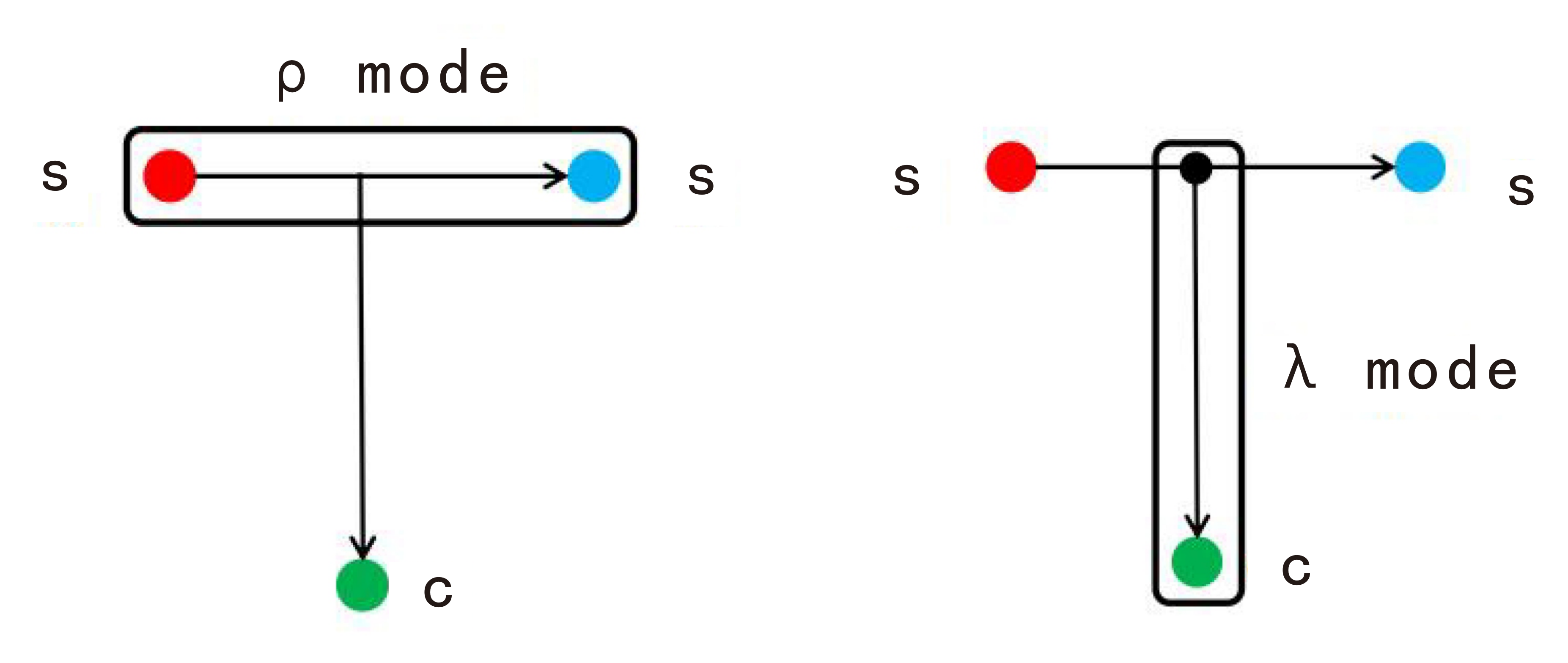} \vspace{-0.1in}
\caption{$\Omega_c$ system with $\lambda-$ or $\rho-$ mode excitations.}\label{mode}
\end{center}
\end{figure}

Within the conventional three-quark picture, a natural interpretation of the five narrow states observed in 2017 is that they constitute the lowest $\lambda$-mode $1P$ multiplet~\cite{Karliner:2017kfm,Chen:2017mug}, i.e., $L_\rho=0$ and $L_\lambda=1$, as given in Fig.~\ref{mode}. Since the two identical strange quarks have a symmetric flavor wave function, the Pauli principle requires the spin of the $ss$ subsystem to be $S_\rho=1$. Coupling $S_\rho$ with $L_\lambda$ gives the total angular momentum of the light degrees of freedom, $j_\ell=0,1,2$. Further coupling to the charm-quark spin produces five states,
\begin{equation}
J^P=\left\{\frac{1}{2}^-,\quad
\left(\frac{1}{2}^-,\frac{3}{2}^-\right),\quad
\left(\frac{3}{2}^-,\frac{5}{2}^-\right)\right\},
\end{equation}
corresponding to $j_\ell=0$, $1$, and $2$, respectively. The correspondence between these five theoretical levels and the five observed peaks makes the complete $1P_\lambda$ multiplet assignment particularly appealing. In increasing-mass order, lattice QCD favors the assignments  $(1/2^-,1/2^-,3/2^-,3/2^-,5/2^-)$
for the five observed states~\cite{Padmanath:2017lng}. Nevertheless, level counting alone is insufficient to determine their individual quantum numbers, and different theoretical studies predict different mass orderings and decay patterns~\cite{Chen:2017mug,Karliner:2017kfm,Wang:2017hej}. A simultaneous description of the mass spectrum and strong decay widths is therefore necessary to establish a consistent assignment of these states. In this context, QCD-inspired quark models provide a widely used phenomenological framework for interpreting newly observed hadronic states~\cite{Godfrey:1985xj,Capstick:1986ter}. Among them, the chiral quark model incorporates Goldstone-boson exchanges between light quarks and has been successfully applied to the description of hadron spectra~\cite{Vijande:2004he,Segovia:2008zza,Segovia:2008zz,Ortega:2016hde}, hadron--hadron interactions~\cite{Fernandez:1993hx,Valcarce:1994nr,Ortega:2016mms,Ortega:2016pgg}, and multiquark systems~\cite{Vijande:2006jf,Yang:2020atz,Huang:2023jec}. In the baryon sector, it has also been employed to investigate the spectra and strong decays of excited $N$ and $\Sigma$ baryons~\cite{Tan:2025kjk,Yao:2025qor,Xiao:2015gra}, providing phenomenological interpretations of states such as $N(1535)$, the Roper resonance $N(1440)$, and $\Sigma(1750)$. These studies highlight the important role of Goldstone-boson exchange in baryon dynamics and motivate the application of the chiral quark model to the $\Omega_c$ spectrum.

Actually, our group have previously investigated the $\Omega_c$ system within the chiral quark model from both molecular and conventional three-quark perspectives~\cite{Huang:2017dwn,Yang:2017rpg}. However, in the previous three-quark calculation, the spin--orbit and tensor interactions were not included, and consequently the fine splittings among the different $J^P$ members of the $\lambda$-mode $1P$ multiplet could not be resolved. Motivated by the successful application of the same framework to the $\Xi_c$ system~\cite{Hu:2025brq}, it is natural to revisit the $\Omega_c$ spectrum within an improved formulation. In particular, employing a common set of model parameters for the $\Xi_c$ and $\Omega_c$ systems provides a more stringent test of the model and enhances its predictive power. Accordingly, in this work we systematically investigate the low-lying $2S$, $1P$, and $1D$ excitations of the $\Omega_c$ baryons. Their masses are calculated within the chiral quark model, while their strong decay widths are evaluated using the $^3P_0$ quark-pair-creation mechanism. Our aim is to provide a simultaneous description of the measured masses and decay widths, clarify the possible assignments of the observed $\Omega_c$ states, and predict the properties of the missing excitations.

This paper is organized as follows. In Sec.~II, we briefly introduce the theoretical framework. Sec.~III presents the numerical results and discussions. Finally, a summary is given in Sec.~IV.

\section{THEORETICAL FRAMEWORK}

As an application of chiral symmetry, the chiral quark model contains kinematic term, confinement, one-gluon exchange, and Goldstone-boson exchange as

\begin{eqnarray}
H & =& \sum_{i=1}^{3}\left( m_i+\frac{p^2_i}{2m_i}\right)-T_{cm} +\sum_{j>i=1}^{3}\left[ V_{CON}(\boldsymbol{r}_{ij}) \right.\nonumber \\
&&\left.+V_{OGE}(\boldsymbol{r}_{ij})+V_{GBE}(\boldsymbol{r}_{ij}) \right] ,
\end{eqnarray}
where $m_i$ is the constituent mass of quark (antiquark), $\bf{p}_i$ is the momentum of quark $i$, $T_{cm}$ is the kinetic energy of the center-of-mass, $\boldsymbol{r}_{ij}$ is the relative coordinate between quark $i$ and $j$, and $V_{CON}$, $V_{OGE}$, $V_{GBE}$ are the color confinement, one-gluon exchange potential, and Goldstone-boson exchange potentials. These three kinds of potentials reveal the most relevant features of QCD in the low energy regime, i.e., color confinement, asymptotic freedom and spontaneous chiral symmetry breaking. The explicit forms of these three potentials are given in Appendix~\ref{appendix-potential} and more details can be found in Ref.\cite{Vijande:2004he,Hu:2025brq}.

Since the $\Omega_c$ system is the SU(3) flavor partner of the $\Xi_c^{\prime}$, to maintain the overall consistency of our study, we directly apply the model parameters determined in our previous study of the $\Xi_c^{\prime}$ system ~\cite{Hu:2025brq} without further adjustment. The model parameters in addition with the fitted ground state hadron mass spectra are presented in the Table~\ref{4-0} and Table~\ref{4-3}, respectively.

\begin{table}[!htb]
\centering
\caption{Quark model parameters used in the calculation, which are exactly the same with our previous work on $\Xi_c^{(\prime)}$ family~\cite{Hu:2025brq}. Here, the numbers with superscript "$\ast$" means they are previously fixed during the fit.\label{4-0}}
\begin{tabular}{c|cc c}
 \hline \hline
                   &$m_{u,d}$ (MeV)   &~~~~313$^\ast$ & \\
  Quark masses      &$m_s$ (MeV)  &~~~~555$^\ast$ & \\
                    &$m_c$ (MeV)  &~~~~1800$^\ast$ & \\  \hline
                   &$\Lambda_\pi$ (fm$^{-1}$)  &~~~~4.20$^\ast$ & \\
                   &$\Lambda_{\eta,K}$ (fm$^{-1}$)      &~~~~5.20$^\ast$ & \\
                   &$m_\pi$ (fm$^{-1}$)  &~~~~0.70$^\ast$ & \\
Goldstone bosons   &$m_K$ (fm$^{-1}$)  &~~~~2.51$^\ast$ & \\
                   &$m_\eta$ (fm$^{-1}$)  &~~~~~2.77$^\ast$ & \\
                   &$g^2_{ch}/(4\pi)$  &~~~~0.54$^\ast$ & \\
                   &$\theta_P(^\circ)$  &~~~~-15$^\ast$ & \\  \hline
                   &$a_c$ (MeV$\cdot$fm$^{-1}$)  &~~~~108.34 & \\
     Confinement
                    &$\Delta$ (MeV)  &~~~-82.99 & \\
                    &$a_{s}$     &~~~~0.777$^\ast$ & \\ \hline
                   &$m_\sigma$ (fm$^{-1}$)  &~~~~3.42$^\ast$ & \\
                   &$\Lambda_\sigma$ (fm$^{-1}$)  &~~~~4.20$^\ast$ & \\
Scalar nonet       &$\Lambda_{a_0,\kappa,f_0}$ (fm$^{-1}$)  &~~~~5.20$^\ast$ & \\
                   &$m_{a_0,\kappa,f_0}$ (fm$^{-1}$)  &~~~~4.97$^\ast$ & \\  \hline
                 &$\hat{r}_0=\hat{r}_g~$(MeV$\cdot$fm)  &~~~~24.7$^\ast$ & \\
                    &$\alpha_{uu}$  &~~~~0.477 & \\
               &$\alpha_{us}$  &~~~~0.566 & \\
                    &$\alpha_{ss}$  &~~~~0.502 & \\
      OGE                &$\alpha_{uc}$ &~~~~0.713 & \\
                    &$\alpha_{sc}$  &~~~~0.668 & \\ 
                     &$\alpha_{u\bar{u}}$  &~~~~0.566 & \\
                      &$\alpha_{u\bar{s}}$  &~~~~0.534 & \\
               &$\alpha_{s\bar{s}}$  &~~~~0.395 & \\
               &$\alpha_{u\bar{c}}$  &~~~~0.626 & \\
                    &$\alpha_{s\bar{c}}$  &~~~~0.530 & \\  \hline \hline
\end{tabular}
\end{table}

\begin{table}[h]
\centering
\caption{The fitted masses of ground state baryons and mesons in units of MeV.\label{4-3}}
\begin{tabular}{ccccccccccccccccccc}
\hline \hline
~     &N~       &$\Delta$~  &$\Lambda$~     &$\Sigma$~          &$\Sigma^{*}$~                                                     \\ \hline
CQM~ &940     &1285   &1119          &1143            &1385                                                                                                                                     \\
Exp.~ &939    &1232      &1116~          &1189~             &1383~                                                                                    \\ \hline

 ~   &$\Xi$~  &$\Xi^{*}$  &$\Omega$~  &$\Lambda_c$~      &$\Sigma_{c}$~        \\ \hline
CQM~    &1350    &1501  &1656~  &2274~    &2475~                          \\
Exp.~   &1318~ &1533  &1672~  &2286~     &2455~                                             \\ \hline

~ &$\Sigma_{c}^{*}$~    &$\Xi_c$~  &$\Xi_c^{\prime}$~ &$\Xi_{c}^{\prime*}$~ &$\Omega_{c}$~                   \\ \hline
CQM~ &2541~         &2473~  &2565~       &2637~    &2689~                  \\
Exp.~   &2520~         &2471~ &2579~    &2645~    &2695~       
\\ \hline
~    &$\Omega_{c}^{*}$   &$\pi$~   &$\rho$~   &$K$~  &$K^{*}$                                 \\ \hline
CQM~  &2765~    &140~     &821~&494~   &975~        \\
Exp.~ &2770~  &140~     &775~        &495~    &892~       \\ \hline

~       &$\omega$~ &$\eta$  &$\eta^{\prime}$~ &$\phi$~ &$D$~              \\ \hline
CQM~      &701~ &508 &958~  &1040~ &1863~        \\      
Exp.~  &782~ & 548 &958~  &1020~ &1864~             
\\ \hline
~         &$D^{*}$~ &$D_s$~    &$D^{*}_s$~            \\ \hline
CQM~      &2063~&1967~     &2203~ \\      
Exp.~  &2007~&1968~     &2112~         
\\ \hline
\hline \hline
\end{tabular}
\end{table}

For the $\Omega_c$ baryon, the two light quarks are identical, so its flavor wave function is
\begin{eqnarray}
    |\Omega_c\rangle_f = ssc,
\end{eqnarray}

Taking into account the antisymmetric color wave function of any baryon as
\begin{eqnarray}
    |\Omega_c\rangle_c = \frac{1}{\sqrt{6}}(rgb + gbr + brg - grb - bgr - rbg),
\end{eqnarray}
under $S-L$ coupling representation, the total wave function of the $\Omega_c$ with total angular momentum $J$ can be represented as
\begin{eqnarray}
    |\Omega_c\rangle_J = |\Omega_c\rangle_c \otimes |\Omega_c\rangle_f \otimes \left[|\Omega_c\rangle_S \otimes |\Omega_c\rangle_L\right]_J,
\end{eqnarray}
where $|\Omega_c\rangle_S$ and $|\Omega_c\rangle_L$ are spin and spatial wave functions that are expanded as
\begin{eqnarray}
    &&|\Omega_c\rangle_S = \left[\left[\psi_{S_1}^{\rm spin}\otimes\psi_{S_2}^{\rm spin}\right]_{S_\rho}\otimes \psi_{S_3}^{\rm spin}\right]_{S},\\
    &&|\Omega_c\rangle_L = \left[\psi_{L_{\rho}}^{\rm orbit}\otimes\psi_{L_{\lambda}}^{\rm orbit}\right]_{L}.
\end{eqnarray}
Here, $S_\rho$ denotes the total spin of the two light quarks in the $\Omega_c$ baryon, $S$ is the total spin, $L_{\rho}$ is the relative orbital angular momentum between the two light quarks, $L_\lambda$ is the relative orbital angular momentum between the charm quark and the light quark pair, and $L$ is the total orbit angular momentum.

For the spatial wave functions, we adopt the Rayleigh-Ritz variational method to solve the eigenvalue problem, where a basis expansion of the trial spatial wave function is performed. In this work, we select the widely used Gaussian expansion method (GEM)~\cite{Hiyama:2003cu} to expand each relative motion in the system. The GEM has proven to be an accurate and universal few-body calculation method~\cite{Hu:2020zwc,Yang:2020fou,Tan:2020cpu}, whose key idea is to expand the radial part of the orbital wave function with a set of Gaussian functions as
\begin{eqnarray}
\psi^{\rm orbit}_l = \sum\limits_{n=1}^{n_{max}} C_n N_{nl}^r e^{-\nu_n r^2}\mathcal{Y}_l(\boldsymbol{r})
\end{eqnarray}
where $\mathcal{Y}_l(\boldsymbol{r})$ is the solid spherical harmonic function, $N_{nl}$ is the normalization constant,
\begin{eqnarray}
\emph{N}_{nl}=\left(\frac{2^{l+2}(2\nu_{n})^{l+3/2}}{\sqrt\pi(2l+1)!!}\right)^{\frac{1}{2}},
\end{eqnarray}
and $C_{n}$ are the variational parameters, which are determined by the dynamics of the system. The Gaussian sizes are chosen as the following geometric progression
\begin{eqnarray}
\nu_{n}=\frac{1}{r^{2}_{n}}, \quad r_{n}=r_{min}\left(\frac{r_{max}}{r_{min}}\right)^{\frac{n-1}{n_{max}-1}},
\end{eqnarray}
where $r_{min}$ and $r_{max}$ are parameters, $n_{max}$ is the number of Gaussian functions, which are determined by the convergence of the results. In this work, we find that after setting $r_{min}=0.1$ fm, $r_{max}=2$ fm, and $n_{max}=8$, the spectra of the low-lying excited $\Omega_c$ baryons are sufficiently convergent.

Finally, to adopt the language of heavy quark symmetry, a representation transformation is performed, which is explicitly expressed as
\begin{eqnarray}
    \left[\psi_{J_l}\otimes\psi_{S_Q}^{\rm spin}\right]_J &=&(-1)^{L+S_\rho+J+\frac{1}{2}} \sum\limits_S\sqrt{2 J_l+1}\nonumber\\
    &&\times\sqrt{2 S+1}\left\{\begin{array}{ccc}
    L & S_\rho & J_l \\
    S_Q & J & S
    \end{array}\right\}\nonumber\\
    &&\times\left[\psi_S^{\rm spin} \otimes \psi_L^{\rm orbit}\right]_J.
\end{eqnarray}
Here, $S_Q$ denotes the spin of the heavy quark, while
$\boldsymbol{J}_l=\boldsymbol{L}+\boldsymbol{S}_\rho$ represents the total
angular momentum of the light degrees of freedom. In the heavy-quark limit,
the spin-dependent interactions involving the heavy quark are suppressed
by the inverse of the heavy-quark mass, and $J_l$ becomes a good quantum
number. Therefore, the $J_l-S_Q$ coupling scheme provides a convenient
classification of singly heavy baryons. It should be emphasized that the
$S-L$ and $J_l-S_Q$ coupling schemes are related by a unitary transformation
and yield identical mass eigenvalues when the same Hamiltonian is
diagonalized in a complete model space.

\subsection{Two body decay mechanism}
To calculate the strong decay widths of the $\Omega_c$ baryons, 
we employ the $^3P_0$ model, which has been widely applied 
and proven to be effective in studying OZI-allowed strong 
decays of hadrons \cite{Micu:1968mk,LeYaouanc:1988fx,Ackleh:1996yt,Roberts:1992esl}. 
According to this model, a quark-antiquark pair with vacuum 
quantum numbers $J^{PC}=0^{++}$ is created from the vacuum. 
Through a quark rearrangement process, this newly created 
$q\bar{q}$ pair recombines with the three constituent quarks 
of the initial baryon, producing the final meson and baryon. 

The transition operator of the $^{3}P_{0}$ model was originally formulated in momentum space. However,when applied to the mass shifts of light mesons, it was found to produce unexpectedly large shifts \cite{Chen:2017mug}. To 
resolve this issue, convergence and damping factors were 
introduced in Refs.~\cite{Chen:2017mug,Huang:2023jec}, which reflect the cutoffs on the momentum and distance, i.e., the created quark pair should not have too large momentum, and the distance from the initial hadron at which it is created should not be too far away from the initial hadron. 
The modified transition operator, along with the detailed 
expressions for the helicity amplitude and the spatial integral, 
is presented in Appendix~\ref{sec:3p0}.

With the modified transition operator, the partial decay width 
for a process $A \rightarrow B + C$ is calculated as
\begin{align}
\Gamma = \pi^{2} \frac{|\mathbf{p}|}{M_{A}^{2}} 
\frac{\mathcal{S}}{2J_A + 1}
\sum_{M_{J_A},M_{J_B},M_{J_C}}
\left|\mathcal{M}^{M_{J_A} M_{J_B} M_{J_C}}\right|^{2}, 
\end{align}
where $\mathcal{M}$ is the helicity amplitude, $\mathcal{S} \equiv 1/(1+\delta_{BC})$ is a statistical factor for identical particles in the final state, and $\boldsymbol{p}$
is the momentum of the final hadron $B$ or $C$ in the 
center-of-mass frame of $A$:
\begin{align}
|\mathbf{p}| = \frac{\sqrt{[M_{A}^{2}-(M_{B}+M_{C})^{2}]
[M_{A}^{2}-(M_{B}-M_{C})^{2}]}}{2M_{A}}. 
\end{align}

The parameters $\gamma$, $f$, and $R_0$ in the transition 
operator are taken from Ref.~\cite{Chen:2017mug}. These values simultaneously reproduce the width of the $\rho \rightarrow \pi\pi$ and avoid the mass shift issue:
\begin{align}
\gamma = 32,\qquad f = 0.5\ \text{fm},\qquad R_0 = 1.0\ \text{fm}. 
\end{align}

\section{Results and discussions}
In this section, we present the mass spectra of the 2$S$, 1$P$, and 1$D$ $\Omega_c$ states obtained within the chiral quark model, as well as the corresponding decay widths under the $^{3}P_{0}$ decay mechanism.

The experimental values of well-established excited $\Omega_c$ baryons from PDG are listed in Table~\ref{1-4}, while our calculated mass spectra for the $\Omega_c$ are presented in Table~\ref{4-2}, along with their corresponding quantum numbers. Here $n_\lambda$, $L_\rho$ and $S_\rho$ denote the radial quantum number, relative orbital angular momentum and the total spin of the two light quarks ($ss$), respectively. $L$ is the total orbital angular momentum of the baryon, $J_l$ is the light degrees of freedom, and $J$ denotes the total angular momentum. In addition, the corresponding decay widths for these states are listed in Table~\ref{width}.

\begin{table*}[ht]
\caption{The basic experimental information of the excited $\Omega_{c}$ baryons  observed so far summarizes the $J^{P}$, mass, total width, and decay mode.\label{1-4}}
\setlength{\tabcolsep}{14pt}
\begin{tabular}{ccccc}
\hline \hline
State                   &$J^{P}$~\cite{ParticleDataGroup:2024cfk}   &Mass (MeV)~\cite{ParticleDataGroup:2024cfk}          &Width (MeV)~\cite{ParticleDataGroup:2024cfk}     &Decay modes~\cite{ParticleDataGroup:2024cfk}                  \\ \hline
$\Omega_{c}(3000)^{0}$ &$?^?$     &$3000.46\pm0.25$             &$3.83\pm0.23$              &$\Xi_{c}^{+}K^-$\\
$\Omega_{c}(3050)^{0}$ &$?^?$     &$3050.17\pm0.19$             &$<$1.8               &$\Xi_{c}^{+}K^-$ \\
$\Omega_{c}(3065)^{0}$ &$?^?$     &$3065.58\pm0.21$             &$3.4^{+0.7}_{-0.8}$       &$\Xi_{c}^{+}K^-$ \\
$\Omega_{c}(3090)^{0}$ &$?^?$     &$3090.15\pm0.26$             &$8.48\pm0.44$       &$\Xi_{c}^{+}K^-$ \\
$\Omega_{c}(3119)^{0}$ &$?^?$     &$3118.98\pm0.12$             &$<$2.5      &$\Xi_{c}^{+}K^-$ \\
$\Omega_{c}(3185)^{0}$ &$?^?$     &$3185\pm1.7$             &$50\pm7$      &$\Xi_{c}^{+}K^-$ \\
$\Omega_{c}(3327)^{0}$ &$?^?$     &$3327.1\pm1.2$             &$20\pm5$      &$\Xi_{c}^{+}K^-$ \\
\hline \hline
\end{tabular}
\end{table*}

\begin{table*}[ht]
\renewcommand{\arraystretch}{1.5}
 \caption{The quantum numbers of $\Omega_{c}$ baryons in the $1S$, $2S$, $1P$ and $1D$ states and the corresponding masses in units of MeV. Here $\Omega_c$, $\widetilde{\Omega}_c$ and $\widehat{\Omega}_c$ denote the $\lambda$-mode, $\rho$-mode and mixed-mode excitations, respectively.\label{4-2}}
 \begin{tabular}{ccccc|ccccc}
 \hline
 \hline
 State~~~~&  Mass~~~~&$L_{\rho}$~~~~ & $L_{\lambda}$~~~~ &[$L$, $S_{\rho}$, $J_{l}$, $J$] ~~~~& State~~~~&  Mass~~~~&$L_{\rho}$~~~~ & $L_{\lambda}$~~~~ &[$L$, $S_{\rho}$, $J_{l}$, $J$] ~~~~\\
 \hline
 $\Omega_{c}(\frac{1}{2}^{+},1S)$&2689   &0&0&[0,~1,~1,~$\frac{1}{2}$]&$\widehat{\Omega}_{c2}(\frac{3}{2}^{+},1D)$&3360  &1&1&[2,~0,~2,~$\frac{3}{2}$]\\ 
 $\Omega_{c}^{*}(\frac{3}{2}^{+},1S)$&2765 &0&0&[0,~1,~1,~$\frac{3}{2}$]&$\widehat{\Omega}_{c2}(\frac{5}{2}^{+},1D)$&3380 &1&1&[2,~0,~2,~$\frac{5}{2}$]\\ 
 $\Omega_{c}(\frac{1}{2}^{+},2S)$& 3168 &0&0&[0,~1,~1,~$\frac{1}{2}$]&$\Omega_{c1}(\frac{1}{2}^{+},1D)$&3253  &0&2&[2,~1,~1,~$\frac{1}{2}$]  \\
$\Omega_{c}(\frac{3}{2}^{+},2S)$& 3219 &0&0&[0,~1,~1,~$\frac{3}{2}$]&$\Omega_{c1}(\frac{3}{2}^{+},1D)$&3266  &0&2&[2,~1,~1,~$\frac{3}{2}$]  \\
$\Omega_{c0}(\frac{1}{2}^{-},1P)$& 2999 &0&1&[1,~1,~0,~$\frac{1}{2}$]&$\Omega_{c2}(\frac{3}{2}^{+},1D)$&3291  &0&2&[2,~1,~2,~$\frac{3}{2}$]  \\
$\Omega_{c1}(\frac{1}{2}^{-},1P)$& 3034 &0&1&[1,~1,~1,~$\frac{1}{2}$]&$\Omega_{c2}(\frac{5}{2}^{+},1D)$&3281  &0&2&[2,~1,~2,~$\frac{5}{2}$]  \\
$\Omega_{c1}(\frac{3}{2}^{-},1P)$& 3042 &0&1&[1,~1,~1,~$\frac{3}{2}$]&$\Omega_{c3}(\frac{5}{2}^{+},1D)$&3296  &0&2&[2,~1,~3,~$\frac{5}{2}$]  \\
$\Omega_{c2}(\frac{3}{2}^{-},1P)$& 3053 &0&1&[1,~1,~2,~$\frac{3}{2}$]&$\Omega_{c3}(\frac{7}{2}^{+},1D)$&3305  &0&2&[2,~1,~3,~$\frac{7}{2}$]  \\
$\Omega_{c2}(\frac{5}{2}^{-},1P)$& 3078 &0&1&[1,~1,~2,~$\frac{5}{2}$]&$\widetilde{\Omega}_{c1}(\frac{1}{2}^{+},1D)$&3366  &2&0&[2,~1,~1,~$\frac{1}{2}$] \\
$\widetilde{\Omega}_{c1}(\frac{1}{2}^{-},1P)$& 3123 &1&0&[1,~0,~1,~$\frac{1}{2}$]&$\widetilde{\Omega}_{c1}(\frac{3}{2}^{+},1D)$&3412  &2&0&[2,~1,~1,~$\frac{3}{2}$]  \\
$\widetilde{\Omega}_{c1}(\frac{3}{2}^{-},1P)$& 3140 &1&0&[1,~0,~1,~$\frac{3}{2}$]&$\widetilde{\Omega}_{c2}(\frac{3}{2}^{+},1D)$&3382  &2&0&[2,~1,~2,~$\frac{3}{2}$]  \\
&&&&&$\widetilde{\Omega}_{c2}(\frac{5}{2}^{+},1D)$&3447  &2&0&[2,~1,~2,~$\frac{5}{2}$] \\
&&&&&$\widetilde{\Omega}_{c3}(\frac{5}{2}^{+},1D)$&3436  &2&0&[2,~1,~3,~$\frac{5}{2}$] \\
&&&&&$\widetilde{\Omega}_{c3}(\frac{7}{2}^{+},1D)$&3470  &2&0&[2,~1,~3,~$\frac{7}{2}$] \\
\hline
 \hline
 \end{tabular}
 \end{table*}

\begin{table*}[ht]
\renewcommand\tabcolsep{0.1cm}
\renewcommand{\arraystretch}{1.5}
\caption{Left: decay widths (in MeV) of the five $\Omega_c$ states observed by LHCb Collaboration~\cite{LHCb:2023sxp} as candidates for the $1P$ or $2S$ state of $\Omega_c$. Right: decay widths (in MeV) of $\lambda$-mode $D$ wave excitation of $\Omega_{c}$. Here, ``$\times$'' means this decay cannot happen due to threshold, and ``$\sim0$'' means the corresponding partial decay width is less than 0.01 MeV. \label{width}}
\begin{tabular}{c|cccc|c|cccccccc}
\toprule[1pt]
\diagbox[width=2.55cm,height=1.4cm,dir=NW]{state}{model}
 & $\Xi_{c}\bar{K}$~ & $\Xi_{c}^{\prime}\bar{K}$~~ & $\Xi_{c}^{*}\bar{K}$~~ & $\Gamma_{\text{total}}$
 & \diagbox[width=2.55cm,height=1.4cm,dir=NW]{state}{model}
 & $\Xi_{c}\bar{K}$~ ~& $\Xi_{c}^{\prime}\bar{K}$~ ~& $\Xi_{c}^{*}\bar{K}$~~ & $\Xi_{c}\bar{K}^{*}$ & $\Xi D$ & $\Xi^* D$ & $\Xi D^*$ & $\Gamma_{\text{total}}$ \\
\midrule[1pt]
$\Omega_{c}(\frac{1}{2}^{+},2S,3168)$ & 8.87 & 10.83 & 1.55  & 21.25
 & $\widehat{\Omega}_{c2}(\frac{3}{2}^{+},1D,3360)$   & $\sim$0 & 3.61 & 4.69 & $\times$  & 0.0 & $\times$ & $\times$ & 8.30 \\
$\Omega_{c}^*(\frac{3}{2}^{+},2S,3219)$  & 3.73 & 1.94 & 9.73 & 15.40
 & $\widehat{\Omega}_{c2}(\frac{5}{2}^{+},1D,3380)$   & $\sim$0 & 3.43 & 8.12 & $\times$ & 0.33 & 0.0 & $\times$ & 11.88 \\
$\Omega_{c0}(\frac{1}{2}^{-},1P,2999)$  & 8.89 & $\times$ &  $\times$ & 8.89
 & $\Omega_{c1}(\frac{1}{2}^{+},1D,3253)$   & 3.41 & 1.20 & 0.67 & $\times$  & 0.72 & $\times$ & $\times$ & 6.00 \\
$\Omega_{c1}(\frac{1}{2}^{-},1P,3034)$  & 13.42 & $\times$  & $\times$ & 13.42
 & $\Omega_{c1}(\frac{3}{2}^{+},1D,3266)$ &   2.15 & 0.36 & 1.49 & $\times$  & 1.06 & $\times$ & $\times$ & 5.06 \\
$\Omega_{c1}(\frac{3}{2}^{-},1P,3042)$   & 0.02 & $\times$ & $\times$ & 0.02
 & $\Omega_{c2}(\frac{3}{2}^{+},1D,3291)$  & 1.78 & 0.35 & 1.61 & $\times$ &  1.08 & $\times$ & $\times$ & 4.82 \\
$\Omega_{c2}(\frac{3}{2}^{-},1P,3053)$   & 0.10 & $\times$  & $\times$ & 0.10
 & $\Omega_{c2}(\frac{5}{2}^{+},1D,3281)$  & 0.06 & 0.30 & 1.77 & $\times$ &  1.02 & $\times$ & $\times$ & 3.15  \\
$\Omega_{c2}(\frac{5}{2}^{-},1P,3078)$   & 0.91 & 0.01 & $\times$ & 0.92
 & $\Omega_{c3}(\frac{5}{2}^{+},1D,3296)$  & 0.26 & 1.30 & 6.46 & $\times$ &  0.01 & $\times$ & $\times$ & 8.03 \\
$\widetilde{\Omega}_{c1}(\frac{1}{2}^{-},1P,3123)$  & $\sim$0 & 12.61  & $\times$ & 12.61
 & $\Omega_{c3}(\frac{7}{2}^{+},1D,3305)$ & 2.14 & 0.38 & 0.51 & $\times$ &  0.07 & $\times$ & $\times$ & 3.10 \\
$\widetilde{\Omega}_{c1}(\frac{3}{2}^{-},1P,3140)$  & $\sim$0 & 0.38 & 7.15 & 7.53
 & $\widetilde{\Omega}_{c1}(\frac{1}{2}^{+},1D,3366)$  & 3.50 & 1.72 & 0.90 & $\times$ & 40.92 & $\times$ & $\times$ & 47.04 \\
 & & & &
 & $\widetilde{\Omega}_{c1}(\frac{3}{2}^{+},1D,3412)$  & 0.72 & 0.15 & 1.39  & $\times$ & 1.34 & 4.95 & $\times$ & 8.55 \\
 & & & &
 & $\widetilde{\Omega}_{c2}(\frac{3}{2}^{+},1D,3382)$  & 1.16 & 0.19 & 1.52  & $\times$ & 2.22 & 0.50 & $\times$ & 5.59 \\
 & & & &
 & $\widetilde{\Omega}_{c2}(\frac{5}{2}^{+},1D,3447)$  & 0.14 & 1.02 & 1.27  & $\times$ & 29.49 & 9.56 & 125.15 & 166.63 \\
 & & & &
 & $\widetilde{\Omega}_{c3}(\frac{5}{2}^{+},1D,3436)$  & 0.15 & 1.05 & 1.57 & $\times$ & 7.83 & 1.76 & 20.41 & 32.77 \\
 & & & &
 & $\widetilde{\Omega}_{c3}(\frac{7}{2}^{+},1D,3470)$  & 3.75 & 1.17 & 2.85 & 0.01 & 47.11 & 6.37 & 5.67 & 66.93 \\
\bottomrule[1pt]
\end{tabular}
\end{table*}

\if
\renewcommand\tabcolsep{0.1cm}
\renewcommand{\arraystretch}{1.7}
\begin{table*}[!htbp]
\centering
\caption{The decay widths (in MeV) of the five $\Omega_c$ states observed by LHCb Collaboration~\cite{LHCb:2023sxp} as candidates for the $1P$ or $2S$ state of $\Omega_c$, where the states are labeled by their $J^P$ quantum numbers, radial and orbital excitations, and masses (in MeV). Here, the symbol "$\times$" means this decay cannot happen due to threshold, and "$\sim0$" means the corresponding partial decay width is less than 0.01 MeV. \label{PPP}}
\begin{tabular}{l|ccccccc}
\toprule[1pt]
\diagbox[width=2.2cm,height=1.0cm,trim=l]{state}{model}
 & $\Xi_{c}\bar{K}$ & $\Xi_{c}^{\prime}\bar{K}$ & $\Xi_{c}^{*}\bar{K}$ &   $\Gamma_{total}$ \\
\midrule[1pt]
$\Omega_{c}(\frac{1}{2}^{+},2S)$ & 8.87 & 10.83 & 1.55  & 21.25 \\
$\Omega_{c}^*(\frac{3}{2}^{+},2S)$  & 3.73 & 1.94 & 9.73 & 15.40 \\
$\Omega_{c0}(\frac{1}{2}^{-},1P)$  & 8.89 & $\times$ &  $\times$ & 8.89  \\
$\Omega_{c1}(\frac{1}{2}^{-},1P)$  & 13.42 & $\times$  & $\times$ &13.42 \\
$\Omega_{c1}(\frac{3}{2}^{-},1P)$   & 0.02 & $\times$ & $\times$ &0.02  \\
$\Omega_{c2}(\frac{3}{2}^{-},1P)$   & 0.10 & $\times$  & $\times$ &0.10  \\
$\Omega_{c2}(\frac{5}{2}^{-},1P)$   & 0.91 & 0.01 & $\times$ &0.92  \\
$\widetilde{\Omega}_{c1}(\frac{1}{2}^{-},1P)$  & 0 & 12.61  & $\times$ & 12.61\\
$\widetilde{\Omega}_{c1}(\frac{3}{2}^{-},1P)$  & 0 & 0.38 & 7.15 & 7.53  \\
\bottomrule[1pt]
\end{tabular}
\end{table*}
\fi

\if
\renewcommand\tabcolsep{0.1cm}
\renewcommand{\arraystretch}{1.7}
\begin{table*}[!htbp]
\centering
\caption{The decay widths (in MeV) of  $\lambda$-mode $D$ wave excitation of $\Omega_{c}$, where the states are labeled by their $J^P$ quantum numbers, radial and orbital excitations, and masses (in MeV). Here, "$\sim0$" means the corresponding partial decay width is less than 0.01 MeV. \label{DDD}}
\begin{tabular}{l|ccccccccccc}
\toprule[1pt]
\diagbox[width=2.2cm,height=1.0cm,trim=l]{state}{model}
 & $\Xi_{c}\bar{K}$ & $\Xi_{c}^{\prime}\bar{K}$ & $\Xi_{c}^{*}\bar{K}$ &$\Xi_{c}\bar{K}^{*}$ &  $\Xi D$ &$\Xi^* D$&$\Xi D^*$ & $\Gamma_{total}$ \\
\midrule[1pt]
$\widehat{\Omega}_{c2}(\frac{3}{2}^{+},1D)$   & $\sim$0 & 3.61 & 4.69 & $\times$  & 0.0 & $\times$ & $\times$ & 8.30 \\
$\widehat{\Omega}_{c2}(\frac{5}{2}^{+},1D)$   & $\sim$0 & 3.43 & 8.12 & $\times$ & 0.33 & 0.0 & $\times$ & 11.88 \\
$\Omega_{c1}(\frac{1}{2}^{+},1D)$   & 3.41 & 1.20 & 0.67 & $\times$  & 0.72 & $\times$ & $\times$ & 6.00 \\
$\Omega_{c1}(\frac{3}{2}^{+},1D)$ &   2.15 & 0.36 & 1.49 & $\times$  & 1.06 & $\times$ & $\times$ & 5.06 \\
$\Omega_{c2}(\frac{3}{2}^{+},1D)$  & 1.78 & 0.35 & 1.61 & $\times$ &  1.08 & $\times$ & $\times$ & 4.82 \\
$\Omega_{c2}(\frac{5}{2}^{+},1D)$  & 0.06 & 0.30 & 1.77 & $\times$ &  1.02 & $\times$ & $\times$ & 3.15  \\
$\Omega_{c3}(\frac{5}{2}^{+},1D)$  & 0.26 & 1.30 & 6.46 & $\times$ &  0.01 & $\times$ & $\times$ & 8.03 \\
$\Omega_{c3}(\frac{7}{2}^{+},1D)$ & 2.14 & 0.38 & 0.51 & $\times$ &  0.07 & $\times$ & $\times$ & 3.10 \\
$\widetilde{\Omega}_{c1}(\frac{1}{2}^{+},1D)$  & 3.50 & 1.72 & 0.90 & $\times$ & 40.92 & $\times$ & $\times$ & 47.04 \\
$\widetilde{\Omega}_{c1}(\frac{3}{2}^{+},1D)$  & 0.72 & 0.15 & 1.39  & $\times$ & 1.34 & 4.95 & $\times$ & 8.55 \\
$\widetilde{\Omega}_{c2}(\frac{3}{2}^{+},1D)$  & 1.16 & 0.19 & 1.52  & $\times$ & 2.22 & 0.50 & $\times$ & 5.59 \\
$\widetilde{\Omega}_{c2}(\frac{5}{2}^{+},1D)$  & 0.14 & 1.02 & 1.27  & $\times$ & 29.49 & 9.56 & 125.15 & 166.63 \\
$\widetilde{\Omega}_{c3}(\frac{5}{2}^{+},1D)$  & 0.15 & 1.05 & 1.57 & $\times$ & 7.83 & 1.76 & 20.41 & 32.77 \\
$\widetilde{\Omega}_{c3}(\frac{7}{2}^{+},1D)$  & 3.75 & 1.17 & 2.85 & 0.01 & 47.11 & 6.37 & 5.67 & 66.93 \\
\bottomrule[1pt]
\end{tabular}
\end{table*}
\fi

\subsection{$\Omega_{c}(3000)$ }
As reported in the latest measurement by the LHCb Collaboration ~\cite{LHCb:2023sxp}, the mass and width of $\Omega_{c}(3000)$ are determined to be
\begin{align}
    \Omega_{c}(3000)^{0}: M &= 3000.44 \pm 0.07~^{+0.07}_{-0.13} \pm 0.23 ~\mathrm{MeV}, \nonumber \\
    \Gamma &=3.83\pm0.23~^{+1.59}_{-0.29} ~\mathrm{MeV}. \nonumber \nonumber
\end{align}
Here the first uncertainty is statistical, the second systematic, and the third (mass only) arises from the uncertainty in the $\Xi_c^+$ mass. However, its quantum numbers have not yet been determined. Numerous theoretical studies have been devoted to the $\Omega_{c}(3000)$, including lattice QCD~\cite{Padmanath:2017lng}, QCD sum rules~\cite{Wang:2017zjw,Wang:2017xam,Agaev:2017lip,Aliev:2017led,Luo:2026elv} and quark models\cite{Ortiz-Pacheco:2020hmj,Wang:2017hej,Weng:2024roa,Zhong:2025oti,Santopinto:2018ljf,Ortiz-Pacheco:2023kjn,Kim:2017jpx}. All of these analyses consistently favor the assignment of $J^P = 1/2^-$ for the $\Omega_{c}(3000)$, identifying it as a $\lambda-$ mode $1P$ state. Nevertheless, a few alternative interpretations have also been proposed. In Ref.\cite{Yang:2021lce}, the QCD sum rule analysis assigns $J^P=1/2^-$ to the $\Omega_{c}(3000)$ as well, but identifies it as a $\rho-$mode $1P$ state. In Ref.\cite{An:2017lwg}, a pentaquark interpretation is proposed for the $\Omega_{c}(3000)$ , also with $J^P=1/2^-$. Furthermore, two scenarios are considered for the $\Omega_{c}(3000)$ by the authors in the framework of diquark model~\cite{Karliner:2017kfm}, (i) In their preferred interpretation, all five observed $\Omega_{c}$ are excitations of the $ss$ diquark with respect to the charm quark, and the $\Omega_{c}(3000)$ is assigned $J^P=1/2^-$. (ii) In an alternative scenario, if the $\Omega_{c}(3090)$ and the $\Omega_{c}(3119)$ are interpreted as $2S$ excitations with positive parity, the $\Omega_{c}(3000)$ would be assigned $J^P=3/2^-$.

From Table \ref{4-2}, our calculated mass of the $\Omega_{c0}(\frac{1}{2}^{-},1P)$ is 2999 MeV, which is in good agreement with the experimental value $3000.44 \pm 0.07$ MeV. However, the predicted decay width from the $^3P_0$ model for this pure state is 8.89 MeV~(see Table~\ref{width}), which is larger than the measured value of $3.83\pm0.23$. This discrepancy can be attributed to the intrinsic uncertainties of the $^3P_0$ model, in particular the quark-pair creation strength $\gamma$ and the spatial wave function overlap, which affect the decay widths at the level of $30\%-50\%$. Considering the mass agreement and the model uncertainty in the decay width, we conclude that the $\Omega_{c}(3000)$ can be interpreted as the $1/2^-$ state of the $\lambda-$mode $1P$ excitation. 

In addition, due to the breaking of heavy-quark symmetry, the $\Omega_{c0}(\frac{1}{2}^{-},1P)$ and $\Omega_{c1}(\frac{1}{2}^{-},1P)$ can mix with each other, which leads to one narrow state and a broader partner~\cite{Wang:2017hej,Zhong:2025oti}. Our calculation yields the mass and width of the $\Omega_{c1}(\frac{1}{2}^{-},1P)$ as 3034MeV and 13.42 MeV, respectively. This suggests that future experiments search for this broader state in the relevant energy region to further test the mixing scenario.

 \subsection{$\Omega_{c}(3050)$, $\Omega_{c}(3065)$ and $\Omega_{c}(3090)$}
 
Based on the measurement reported by the LHCb Collaboration~\cite{LHCb:2023sxp}, the masses and widths of $\Omega_{c}(3050)$, $\Omega_{c}(3065)$, and $\Omega_{c}(3090)$ are determined to be

\begin{align}
    \Omega_{c}(3050)^{0}: M &= 3050.18 \pm 0.04~^{+0.06}_{0.07} \pm 0.23 ~\mathrm{MeV}, \nonumber \\ \nonumber
    \Gamma &=0.67\pm0.17~^{+0.64}_{-0.72} ~\mathrm{MeV}; \\\nonumber
      \Omega_{c}(3065)^{0}: M &= 3065.63 \pm 0.06~^{+0.06}_{-0.06} \pm 0.23 ~\mathrm{MeV}, \\\nonumber 
    \Gamma &=3.79\pm0.20~^{+0.38}_{-0.47} ~\mathrm{MeV}; \\\nonumber 
   \Omega_{c}(3090)^{0}: M &= 3090.16 \pm 0.11~^{+0.06}_{0.10} \pm 0.23 ~\mathrm{MeV}, \nonumber \\  \nonumber
    \Gamma &=8.48\pm0.44~^{+0.61}_{-1.62} ~\mathrm{MeV}.\nonumber
\end{align}
For the $\Omega_{c}(3050)$ and $\Omega_{c}(3065)$, various quantum-number assignments have been proposed. Many analyses favor identifying both states as $1P$ wave excitations with negative parity, predominantly assigning  $J^P=3/2^-$ for $\Omega_{c}(3065)$~\cite{Agaev:2017lip,Wang:2017xam,Zhao:2017fov,Aliev:2017led}, while in Ref.~\cite{Karliner:2017kfm} the authors suggest that $\Omega_{c}(3065)$ could instead be $J^P=5/2^-$within the heavy quark-diquark picture. Alternatively, results from the QCD sum rules that incorporate a relative $P$ wave between the two strange quarks assign $\Omega_{c}(3065)$ as $J^P==1/2^-$~\cite{Wang:2017zjw}, and another theoretical analysis has also considered the $J^P=1/2^-$ possibility for $\Omega_{c}(3050)$, but only after the configuration mixing~\cite{Wang:2017hej}. In addition, some interpretations place these two states within the $2S$ radial excitation scenario~\cite{Agaev:2017lip,Yang:2021lce}.
It should be noted that helicity angle measurements by the LHCb Collaboration disfavor the $J=1/2$ hypothesis for both $\Omega_{c}(3050)$ and $\Omega_{c}(3065)$ with significance levels for excluding this hypothes of 2.2$\sigma$ and 3.6$\sigma$, respectively~\cite{LHCb:2021ptx}.

For the $\Omega_{c}(3090)$, results from Lattice QCD~\cite{Padmanath:2017lng} and QCD sum rules~\cite{Wang:2017xam} have strongly suggested its $J^P=3/2^-$, which is consistent with the assignment from the heavy quark-diquark picture~\cite{Karliner:2017kfm}. Meanwhile, $^3P_0$ model analyses support its interpretation as a $1P$ wave state with either $J^P=3/2^-$ or $5/2^-$~\cite{Zhao:2017fov}. On the other hand, alternative theoretical approaches propose a positive-parity interpretation for $\Omega_{c}(3090)$,  interpreting it as a $2S$ excitation with $J^P=1/2^+$~\cite{Agaev:2017lip}. Finally, It is worth noting that the conclusions of Refs.~\cite{Wang:2017xam,Karliner:2017kfm} do not rule out the possibility of $\Omega_{c}(3090)$ as a $2S$ excitation.

(1) $\mathbf{\Omega_{c}(3050)}$ and $\mathbf{\Omega_{c}(3065)}$: Considering that the helicity analysis has excluded the $J=1/2$ hypothesis for both the
$\Omega_{c}(3050)$ and $\Omega_{c}(3065)$, their spins are constrained to be $3/2$ or higher. Among the $\lambda-$mode  $1P$ excitations in our calculation, the two $J^P=3/2^-$ states are calculated to be 3042 MeV and 3053 MeV,  which are in good agreement with the experimental masses of 3050.18 MeV and 3065.63 MeV, respectively. While in terms of decay properties, the partial widths to $\Xi_c \bar{K}$ for these two states are 0.02 MeV and 0.1 MeV, respectively, both smaller than the corresponding experimental total widths of 0.67 MeV and 3.79 MeV. In the absence of measured branching fractions for individual decay channels, a more detailed comparison is not possible at present. Moreover, it is known that the $^3P_0$ model has limited accuracy in predicting absolute decay widths~\cite{Capstick:2000qj,Santopinto:2018ljf}. 
Based on the overall mass agreement and the consistency of the partial widths with the experimental upper limits, we identify the $\Omega_{c}(3050)$ and $\Omega_{c}(3065)$ as the two $3/2^-$ states of the $\lambda-$model $1P$ excitations.

(2) $\mathbf{\Omega_{c}(3090)}$: For the $\Omega_{c}(3090)$, the $\Omega_{c2}(\frac{5}{2}^{-})$ is calculated to be 3078 MeV, which is close to the experimental mass of 3090.15 MeV. Although the deviation is about 12 MeV, it remains within the typical accuracy of the chiral quark model. The partial width to $\Xi_c \bar{K}$ is 0.91 MeV, smaller than the experimental total width of 8.48 MeV, consistent with the pattern observed for $\Omega_{c}(3050)$ and $\Omega_{c}(3065)$. We therefore assign the $\Omega_{c2}(\frac{5}{2}^{-})$ as a candidate for $\Omega_{c}(3090)$.

 \subsection{$\Omega_{c}(3119)$}
 
 The latest  measurement by the LHCb Collaboration gives the mass and width of $\Omega_{c}(3119)$  as~\cite{LHCb:2023sxp}
\begin{align}
    \Omega_{c}(3119)^{0}: M &= 3118.98 \pm 0.12~^{+0.09}_{-0.23} \pm 0.23 ~\mathrm{MeV}, \nonumber \\
    \Gamma &=0.60\pm0.63~^{+0.90}_{-1.05} ~\mathrm{MeV}. \nonumber \nonumber
\end{align}

The $\Omega_{c}(3119)$ has been extensively investigated, yet its quantum numbers remain controversial. Within the conventional three-quark picture, in Refs.~\cite{Li:2024zze,Weng:2024roa}, the authors tend to identify it as a $1P$ wave $\rho-$mode excitation with
$J^P=3/2^-$ in the framework of relativized quark models. Lattice QCD and some phenomenological quark models analyses favor a $1P$ wave $\lambda$-mode excitation with $J^P=5/2^-$\cite{Padmanath:2017lng,Karliner:2017kfm,Ortiz-Pacheco:2020hmj}. Alternatively, several QCD sum rule studies interpret it as a $2S$ radially excited state with $J^P=3/2^+$\cite{Wang:2017xam,Agaev:2017lip}, and a $^3P_0$ model analysis suggests a $1D$ wave interpretation with $J^P=5/2^+$ or $7/2^+$~\cite{Zhao:2017fov}. 
On the other hand, , the possibility of the molecular and pentaquark interpretations for the $\Omega_{c}(3119)$ cannot be excluded. In Refs.~\cite{Santopinto:2018ljf,Debastiani:2017ewu,Tang:2026uls}, the 
$\Omega_{c}(3119)$ has been interpreted as a $\Xi_c^*\bar{K}$ molecule with $J^P=3/2^-$, while the authors of Ref~\cite{Huang:2017dwn} identified it as a $\Xi D$ molecular resonance with $J^P=1/2^-$. In addition, a pentaquark interpretation has also been proposed. The chiral quark-soliton model interprets it as a $3/2^+$ pentaquark~\cite{Kim:2017jpx,Kim:2017khv,Kim:2018cku}  while constituent quark model analyses assign $J^P=1/2^-$ to it~\cite{An:2017lwg}.

 Our calculations yield two $\rho$-mode $1P$ states, $\tilde{\Omega}_{c1}(1/2^-,1P)$ and $\tilde{\Omega}_{c1}(3/2^-,1P)$, with masses of 3123 MeV and 3140 MeV, respectively. These masses are close to the experimental value of 3118.98 MeV. However, their predicted decay widths are strictly zero within the $^3P_0$ model, which is incompatible with a conventional three-quark interpretation. The vicinity of the $\Xi^*\bar{K}$ and $\Xi D$ thresholds suggests that a molecular or pentaquark interpretation for the $\Omega_{c}(3119)$ is reasonable~\cite{Santopinto:2018ljf,Debastiani:2017ewu,Tang:2026uls,Huang:2017dwn}. Meanwhile, for $\Omega_{c}(3119)$,the possibility of three-quark–pentaquark mixing cannot be ruled out, and a systematic investigation of this scenario is an important direction for future work. In addition, based on our calculations, $\Xi_c^{\prime}\bar{K}$ and $\Xi_c^*\bar{K}$ channels could be useful for future experimental searches for these two $\rho-$mode $1P$ states, as their partial widths are found to be relatively broad.

\subsection{$\Omega_{c}(3185)$ and $\Omega_{c}(3327)$}
 
For the $\Omega_{c}(3185)$, two main interpretations have been proposed so far. Within the conventional three-quark framework, the most widely favored assignment is that of a $2S$ radially excited state with $J^P=1/2^+$ or $3/2^+$~\cite{Karliner:2023okv,Pan:2023hwt,Jakhad:2023mni,Kucukyilmaz:2025rsd,Zhong:2025oti,Wang:2023wii}, although alternative possibilities such as a $1P$ wave $\rho$-mode excitation have also been proposed~\cite{Kucukyilmaz:2025rsd}. On the other hand, exotic interpretations cannot be ruled out. Its mass, lying below the threshold of $\Xi D$, naturally supports a molecular scenario, with several studies identifying the $\Omega_{c}(3185)$ as an $S$ wave $\Xi D$ molecular state with $J^P=1/2^-$ or $3/2^-$~\cite{Feng:2023ixl,Yan:2023ttx,Ozdem:2023okg}. 
 
For the $\Omega_{c}(3327)$, in the conventional three-quark framework, a $D$ wave assignment with $J^P=3/2^+$ or $5/2^+$ has received  support from both non-relativistic potential model~\cite{Luo:2023sra}  and semi-relativistic
constituent quark potential model~\cite{Zhong:2025oti}, while QCD sum rules analyses find consistency with a $\Sigma$-type $D$ wave state with $J^P=1/2^+$, $3/2^+$ or $5/2^+$~\cite{Wang:2023wii}. Another conventional interpretation assigns it as a $2S$ radially excited state~\cite{Karliner:2023okv}. Moreover, the mass of the $\Omega_{c}(3327)$ is close to the threshold of $\Xi D^*$, which naturally motivates molecular interpretations. Several studies have identified it as an $\Xi D^*$ molecular state with $J^P=3/2^-$~\cite{Yan:2023ttx,Feng:2023ixl}, and this molecular interpretation is further supported by predictions of its magnetic dipole moments~\cite{Ozdem:2023okg}.

(1) $\mathbf{\Omega_{c}(3185)}$: Our calculated masses of the two $2S$ states with $J^P=1/2^+$ and $3/2^+$ are 3168 MeV and 3219 MeV, respectively, both of which are close to the experimental value of 3185.1 MeV. The experimental width of the
$\Omega_{c}(3185)$ is $50\pm7$ MeV, whereas the partial widths of these two states to
$\Xi_c\bar{K}$ obtained in our calculation are 8.87 MeV and 3.73 MeV, respectively, both considerably smaller than the experimental width. Considering that our results in the $P$-wave sector also exhibit a similar behavior that the theoretical widths are much smaller, we regard such a deviation as a feature of the present model and therefore consider both $2S$ states as possible candidates for the $\Omega_{c}(3185)$, primarily based on the mass agreement. As for which one corresponds to the $\Omega_{c}(3185)$, future measurements of the $\Xi_c^{\prime}\bar{K}$ and $\Xi_c^*\bar{K}$ channels are required.

(2) $\mathbf{\Omega_{c}(3327)}$: Our calculated masses of the $\widehat{\Omega}_{c2}(\frac{3}{2}^{+}$) and $\Omega_{c3}(\frac{7}{2}^{+})$ states are 3360 MeV and 3305 MeV, respectively, both in reasonable agreement with the experimental value of 3327.1 MeV. The experimentally measured total width of the $\Omega_{c}(3327)$ is $20\pm5$ MeV, while the partial widths of these two states to $\Xi_c\bar{K}$ obtained as 5.60 MeV and 10.11 MeV, respectively, which are again lower than the experimental measurement. Similar in the $\Omega_{c}(3185)$ case, such an underestimation appears to be a general feature of the present calculation.
However, it should be noted here that the partial decay width of the $\widehat{\Omega}_{c2}(\frac{3}{2}^{+}$) state to $\Xi_c\bar{K}$ is nearly zero, which is incompatible with the experimental observation of the $\Omega_{c}(3327)$ in this channel. Based on the above analysis, we exclude the $\widehat{\Omega}_{c2}(\frac{3}{2}^{+}$) assignment for the $\Omega_{c}(3327)$ and identify it instead as the possible $\Omega_{c3}(\frac{7}{2}^{+})$ state.

\subsection{Other $\lambda-$mode $D$ wave $\Omega_{c}$ states}
As predictions, we also study the $D$ wave excitation of $\Omega_{c}$, including pure $\lambda-$mode excitation, pure $\rho-$mode excitation, and mixed ones, whose spectra and decay properties are given in Table~\ref{4-2} and Table~\ref{width}. Since pure $\lambda-$mode excitation may be easier to be measured by experiments~\cite{LHCb:2024eyx}, along with this mode is more similar to the heavy-light mesons, in this subsection we only breifly discuss the behavior of pure $\lambda-$mode excited $\Omega_{c}$ states. As shown in Table~\ref{4-2}, their masses range from approximately 3253 to 3305 MeV. In addition, their decay widths are typically of order a few MeV. For the $\Omega_{c1}(\frac{1}{2}^{+})$ and $\Omega_{c3}(\frac{7}{2}^{+})$, the dominant decay channel is $\Xi_c\bar{K}$. The $\Omega_{c1}(\frac{3}{2}^{+})$ and $\Omega_{c2}(\frac{3}{2}^{+})$ states decay mainly via $\Xi_c\bar{K}$, $\Xi_c^*\bar{K}$ and $\Xi D$, while the $\Omega_{c2}(\frac{5}{2}^{+})$ state decays through $\Xi_c^*\bar{K}$ and $\Xi D$, and the $\Omega_{c3}(\frac{5}{2}^{+})$ state mainly via $\Xi_c^*\bar{K}$. Although their mass positions are close to each other, their dominant decay modes are quite distinct. We hope that our results will help the experiment to identify these missing D wave excitations in the $\Omega_c$ spectrum, and provide a test of the present model.

\section{summary}
Due to its unique quark content, the $\Omega_c$ system—composed of a charm quark and two strange quarks—provides a simplified yet compelling platform for investigating heavy-quark symmetry and the mechanisms of chiral symmetry breaking. Its flavor-singlet nature under $SU(3)_F$ naturally suppresses light-quark mixing and streamlines the excitation spectrum, providing a key opportunity for testing QCD-inspired models and exploring explicit chiral dynamics via Goldstone-boson exchange.

After the discovery of the two ground states, discovered in 2009~\cite{Biagi:1984mu,Solovieva:2008fw}, no significant progress was made in the $\Omega_c$ sector until 2017, when the LHCb and Belle Collaborations achieved a series of breakthroughs in the search for excited $\Omega_c$ baryons~\cite{LHCb:2017uwr,Belle:2017ext,Belle:2017szm,LHCb:2021ptx,LHCb:2023sxp}. In particular, in Ref.~\cite{LHCb:2021ptx}, the measurement by the LHCb Collaboration provided valuable insights by ruling out the $J=1/2$ hypothesis for $\Omega_c(3050)^0$ and $\Omega_c(3065)^0$. Inspired by these experimental advances, we adopt these two states as reference points to constrain the model parameters in our study of the $\Omega_c$ spectrum. In this work, we investigate the mass spectrum of $\Omega_c$ baryons  within the framework of the chiral quark model and the 
$^{3}P_{0}$ decay mechanism, with the help of the Gaussian expansion method. By comparing our calculated mass spectra and decay widths with the available experimental data, we provide the following interpretations for their quantum numbers,

(1) $\Omega_{c}(3000)^0$, $\Omega_{c}(3050)^0$, $\Omega_{c}(3065)^0$ and $\Omega_{c}(3090)^0$ are well described as $\lambda-$mode $1P$ states with $J^P=1/2^-$, $3/2^-$, $3/2^-$, and $5/2^-$, respectively. Their calculated masses are in good agreement with the experimental values, with deviations well within the typical accuracy of the chiral quark model.

(2) $\Omega_{c}(3119)^0$ can not be interpreted as a pure three-quark excitation. Although the $\widetilde{\Omega}_{c1}(\frac{1}{2}^{-},1P)$ and $\widetilde{\Omega}_{c1}(\frac{3}{2}^{-},1P)$ have masses close to the experimental value, their decay widths rule them out as candidates for $\Omega_{c}(3119)$. This suggest that a molecular, pentaquark interpretation, or mixing scenario could be more appropriate for this state.

(3) $\Omega_{c}(3185)^0$ and $\Omega_{c}(3327)^0$ are assigned to the $2S$ and 
$1D$ states, respectively, primarily based on the mass agreement, and the calculated decay widths are also not conflict with such arrangements. For the $\Omega_{c}(3185)^0$, both $J^P=1/2^+$ and $3/2^+$ remain possible, and future measurements of the $\Xi_c^{\prime}\bar{K}$ and $\Xi_c^{*}\bar{K}$ channels are required to distinguish between them. For the $\Omega_{c}(3327)^0$, we tend to identify it as the $\lambda-$mode $D$-wave excitation with $J^P=7/2^+$.

(4) Other $\lambda-$mode D wave $\Omega_{c}$ states are predicted with masses around 3253–3305 MeV, with their dominant decay channels identified, providing useful references for future experimental searches.

We therefore expect that future experimental efforts will test our predictions. Such measurements would not only help clarify the properties of the currently unresolved $\Omega_c$ states, but also deepen our understanding of symmetry principles, strong interaction dynamics, and few-body systems.

\if
In this work, we only use one unified set of parameters to investigate the mass spectra and decay widths, which means the unquenched effect is not taken into account. Although our results are slightly different from the experimental values, but they are qualitatively consistent. However, we cannot jump to conclusions in this paper. A mixed system of three-quark baryon and pentaquark cannot be ignored. Therefore, the study of $\Xi_c$ in the framework of the unquenched quark model, including the higher Fock components with more future experimental and theoretical data is our future work.
\fi

\section*{Acknowledgments}
This work is supported partly by the National Natural Science Foundation of China under Grant Nos. 12305087, 12205249, 12247101, 12475080, the General Project of Natural Science Foundation of colleges and universities of Jiangsu Province under Grant No. 24KJB140001, the Funding for the Qinglan Project of Jiangsu Province.

\section*{Appendix}
\subsection{Explicit potentials}
\label{appendix-potential}

The potentials within the chiral quark model contain color confinement, one-gluon exchange, and Goldstone-boson exchanges. For color confinement potential, we adopt the linear form, which refers to the well-known Cornell potential as
\begin{eqnarray}
\setlength{\abovedisplayskip}{5pt}
\setlength{\belowdisplayskip}{5pt}
V^C_{CON}({{\bf r}_{ij}})  &=&  \boldsymbol{\lambda}_i^c\cdot \boldsymbol{\lambda}_j^c
   (-a_c r_{ij}-\Delta), \\
V^{SO}_{CON}({{\bf r}_{ij}})&=& -(\boldsymbol{\lambda}_i^c \cdot\boldsymbol{\lambda}_j^c )
 \frac{a_{c}r_{ij}}{4m_{i}^{2}m_{j}^{2}}
 [((m_{i}^{2}+m_{j}^{2})(1-2a_{s})\nonumber \\
&&+4m_{i}m_{j}(1-a_{s}))(\boldsymbol{S}_{+}\cdot\boldsymbol{L})\nonumber \\
&&+(m_{j}^{2}-m_{i}^{2})(1-2a_{s})(\boldsymbol{S}_{-}\cdot\boldsymbol{L})].
\end{eqnarray}
Here, $V^{SO}_{CON}({{\bf r}_{ij}})$ denotes the Thomas-precession effect, $a_c$ and $\Delta$ are model parameters, $\boldsymbol{S}_{\pm}=\boldsymbol{S}_{i}\pm\boldsymbol{S}_{j}$, and $\boldsymbol{\lambda}^c$ represents the SU(3) Gell-Mann matrices.

For one-gluon exchange interaction, it contains the so-called coulomb and chromomagnetic interactions, in addition with spin-orbit and tensor potentials, which arise from QCD perturbation effects and low-order relativistic corrections, as
\begin{eqnarray}
V^{C}_{OGE}({{\bf r}_{ij}})&=&  \frac{\alpha_s}{4} \boldsymbol{\lambda}_i^c \cdot \boldsymbol{\lambda}_j^c
   \left[ \frac{1}{r_{ij}}-\frac{\boldsymbol{\sigma}_i\cdot \boldsymbol{\sigma}_j}{6m_im_j}  \frac{e^{-r_{ij}/r_0(\mu)}}{r_{ij}r^2_0(\mu)}\right] ,      \\
V_{OGE}^{SO}({{\bf r}_{ij}})&=&  -\frac{\alpha_s}{16m_i^2m_j^2} \boldsymbol{\lambda}_i^c \cdot \boldsymbol{\lambda}_j^c \left[ \frac{1}{r_{ij}^3}- \frac{e^{-r_{ij}/r_g(\mu)}}{r_{ij}^3} \right.  \nonumber\\
&&\left.\times(1+\frac{r_{ij}}{r_g(\mu)} )\right]
    \times\left[((m_i+m_j)^2 +2m_im_j)\right.\nonumber\\
    &&\left.\times\left(\mathbf{S}_+ \cdot \mathbf{L}\right)  
    +(m_j^2-m_i^2)(\mathbf{S}_- \cdot \mathbf{L} )\right],  \\
V_{OGE}^T({{\bf r}_{ij}}) &=& -\frac{\alpha_s}{16 m_im_j} \boldsymbol{\lambda}_i^c \cdot \boldsymbol{\lambda}_j^c \left[ \frac{1}{r_{ij}^3}- \frac{e^{-r_{ij}/r_g(\mu)}}{r_{ij}}\right.\nonumber\\
&&\left.\times\left(\frac{1}{r_{ij}^2} +\frac{1}{3r_{g}^2(\mu)}+\frac{1}{r_{ij}r_g(\mu)}\right) \right]S_{ij}.
\end{eqnarray}
Here, $\mu$ is the reduced mass of two interacting quarks, $\boldsymbol{\sigma}$ represents the SU(2) Pauli matrices, $S_{ij}$ is the tensor operator, $r_0(\mu)\equiv\hat{r}_0/\mu$, $r_{g}(\mu)\equiv\hat{r}_g/\mu$, with $\hat{r}_0$ and $\hat{r}_g$ being parameters, and $\alpha_s$ denotes the effective flavor-dependent strong coupling constant.

For Goldstone-boson exchange potentials, the expressions are 
\begin{eqnarray}
V_{GBE}({{\bf r}_{ij}})&=&V_{\pi}({{\bf r}_{ij}})+V_{K}({{\bf r}_{ij}})+V_{\eta}({{\bf r}_{ij}})+V_{sc}({{\bf r}_{ij}}), \\
V_{\pi}({{\bf r}_{ij}})&=&\frac{g^2_{ch}}{4\pi}\frac{m^2_\pi}{12m_im_j}\frac{\Lambda^2_\pi m_\pi}{\Lambda^2_\pi-m^2_\pi}\sum_{a=1}^{3}\lambda_i^a \lambda_j^a \nonumber \\
&&\times\left\{(\boldsymbol{\sigma}_i \cdot \boldsymbol{\sigma}_j)\left[ Y(m_{\pi}r_{ij})- \frac{\Lambda^3_\pi}{m^3_\pi}Y (\Lambda_{\pi}r_{ij})\right]\right. \nonumber  \\
&&\left.+\left[H(m_{\pi}{\bf r}_{ij})-\frac{\Lambda^3_\pi}{m^3_\pi} H(\Lambda_{\pi}r_{ij})\right]S_{ij} \right\},    \\
V_{K}({{\bf r}_{ij}})&=&\frac{g^2_{ch}}{4\pi}\frac{m^2_K}{12m_im_j}\frac{\Lambda^2_K m_K}{\Lambda^2_K-m^2_K}\sum_{a=4}^{7}\lambda_i^a \lambda_j^a\nonumber \\
&&\times\left\{(\boldsymbol{\sigma}_i \cdot \boldsymbol{\sigma}_j)\left[ Y(m_{K}r_{ij}) -\frac{\Lambda^3_K}{m^3_K}Y (\Lambda_{K}r_{ij}) \right]  \right.  \nonumber\\
&&\left.+\left[H(m_{K}{\bf r}_{ij})-\frac{\Lambda^3_K}{m^3_K} H(\Lambda_{K}r_{ij})\right] S_{ij} \right\},    \\
V_{\eta}({{\bf r}_{ij}})&=&\frac{g^2_{ch}}{4\pi}\frac{m^2_\eta}{12m_im_j}\frac{\Lambda^2_\eta m_\eta}{\Lambda^2_\eta-m^2_\eta}\nonumber \\
&&\times\left[\cos\theta_{P}(\lambda_i^8 \lambda_j^8)-\sin\theta_{P}(\lambda_i^0 \lambda_j^0)\right]\nonumber \\
&&\times\left\{(\boldsymbol{\sigma}_i \cdot \boldsymbol{\sigma}_j)\left[ Y(m_{\eta}r_{ij})-\frac{\Lambda^3_\eta}{m^3_\eta}Y (\Lambda_{\eta}r_{ij}) \right]\right.\nonumber \\
&&\left.+\left[H(m_{\eta}{\bf r}_{ij})-\frac{\Lambda^3_\eta}{m^3_\eta} H(\Lambda_{\eta}r_{ij})\right] S_{ij} \right\}, \\
Y(x)&=&e^{-x}/x,\quad H(x)=\left(1+\frac{3}{x}+\frac{3}{x^2}\right)Y(x),
\end{eqnarray}
where $\lambda^a$ are the Gell-Mann matrices, $\Lambda_\chi~(\chi=\pi,K,\eta)$ are the cut-offs, $m_{\chi}$ are the masses of Goldstone bosons, and $g^2_{ch}$ is the chiral field coupling constant, which is determined from the $NN\pi$ interaction through
\begin{eqnarray}
\frac{g^2_{ch}}{4\pi}=\frac{9}{25}\frac{g^2_{\pi NN}}{4\pi}\frac{m^2_{u,d}}{m^2_N}.
\end{eqnarray}
Additionally, as an extension of the $\sigma$ meson exchange, the scalar nonet exchange $V_{sc}$ is also included as
\begin{eqnarray}
V_{sc}({{\bf r}_{ij}})&=&V_{a_0}({{\bf r}_{ij}})\sum_{a=1}^{3}
	\lambda_i^a \lambda_j^a+V_{\kappa}({{\bf r}_{ij}})\sum_{a=4}^{7}
	\lambda_i^a \lambda_j^a \nonumber \\
&&+ V_{f_0}({{\bf r}_{ij}})
	\lambda_i^8 \lambda_j^8+V_{\sigma}({{\bf r}_{ij}})\lambda_i^0 \lambda_j^0,   \\
V_{s}({{\bf r}_{ij}}) & =& -\frac{g^2_{ch}}{4\pi} \frac{\Lambda^2_s m_s}{\Lambda^2_s-m^2_s}
	\left[ Y(m_{s}r_{ij})-\frac{\Lambda_s}{m_s}Y(\Lambda_{s}r_{ij})\right] \nonumber  \\
&&+\frac{m^3_s}{2m_im_j}\left[G(m_{s}{r}_{ij})-\frac{\Lambda_{s}^{3}}{m_{s}^{3}}G(\Lambda_{s}{r}_{ij})\right]\nonumber\\
&&\times\mathbf{L}\cdot \mathbf{S}~~(s=a_0,\kappa,f_0,\sigma),\\
G(x)&=&\left(1+\frac{1}{x}\right)\frac{Y(x)}{x},
\end{eqnarray}
with $m_s$ and $\Lambda_s$ are the mass and cutoff that correspond to the scalar meson $s$, respectively.

\subsection{The $^{3}P_{0}$ model}
\label{sec:3p0}
The original transition operator in the momentum 
representation is expressed as~\cite{Chen:2024ukv}
\begin{align}
T &= -3\gamma\sum_m\langle 1m;1-m|00\rangle
\int d\mathbf{p}_4 d\mathbf{p}_5 \, 
\delta(\mathbf{p}_4+\mathbf{p}_5) \notag \\
&\quad \times \mathcal{Y}^m_1\left(\frac{\mathbf{p}_4-\mathbf{p}_5}{2}\right)
\chi_{45}^{1-m}\phi_{45}^0\omega_{45}^0
b^\dagger_4(\mathbf{p}_4)d^\dagger_5(\mathbf{p}_5), \tag{B1}
\end{align}
where $\gamma$ is the quark-pair creation strength, 
$\mathcal{Y}^m_1 = |\mathbf{p}| Y_1^m(\theta_p, \phi_p)$ is 
the solid harmonic polynomial reflecting the $P$-wave 
distribution of the created quark pair, and $\phi_{45}^0$, 
$\omega_{45}^0$, $\chi_{45}^{1-m}$ are the flavor, color, 
and spin wave functions of the created quark pair, 
respectively.

To avoid the large mass shifts encountered in light meson 
decays \cite{Chen:2017mug}, the modified transition 
operator is adopted in coordinate space \cite{Chen:2017mug}:
\begin{align}
T &= -3\gamma\sum_{m}\langle 1m;1-m|00\rangle
\int d\mathbf{r}_4 d\mathbf{r}_5 
\left(\frac{1}{2\pi}\right)^{3/2} 2^{-5/2} f^{-5} \notag \\
&\quad \times \mathcal{Y}_{1}^{m}(\mathbf{r})
e^{-\frac{\mathbf{r}^2}{4f^2}} e^{-\frac{R_{AV}^2}{R_0^2}}
\chi^{1-m}_{45} \phi^{0}_{45} \omega^{0}_{45}
b_4^{\dagger}(\mathbf{r}_4)d_5^{\dagger}(\mathbf{r}_5), \tag{B2}
\end{align}
where $\mathbf{r} = \mathbf{r}_4 - \mathbf{r}_5$ is the relative 
coordinate of the created quark pair, $\mathbf{r}_4$ and 
$\mathbf{r}_5$ are the coordinates of the created quark and 
antiquark, respectively, and $\mathbf{R}_{AV}$ is the relative 
coordinate between the initial baryon $A$ and the created 
quark pair. The exponential factors serve as convergence 
and damping factors, respectively.

With the modified operator, the matrix element for the transition $A \rightarrow B + C$ 
can be written as
\begin{align}
\langle BC|T|A\rangle = \delta^3(\mathbf{P}_A - \mathbf{P}_B - 
\mathbf{P}_C) \mathcal{M}^{M_{J_A}M_{J_B}M_{J_C}}, \tag{B3}
\end{align}
where $\mathbf{P}_B$ and $\mathbf{P}_C$ are the momenta of 
the final-state hadrons in the center-of-mass frame of $A$, 
and $\mathcal{M}$ is the helicity amplitude:
\begin{align}
&\mathcal{M}^{M_{J_A} M_{J_B} M_{J_C}}(A \rightarrow BC) \notag \\
&= \sqrt{8 E_A E_B E_C} 
\left\langle \varphi_C \varphi_B 
\middle| \varphi_A \varphi_0 \right\rangle \sum_{\substack{M_{L_A}, M_{S_A}, \\ M_{L_B}, M_{S_B}, \\ 
M_{L_C}, M_{S_C}, m}} 
I_{M_{L_B}, M_{L_C}}^{M_{L_A}, m} \notag \\
&\quad \times  \langle 1m\,1{-m}|00\rangle \left\langle \chi_{S_C M_{S_C}}^{235} 
\chi_{S_B M_{S_B}}^{14} \middle| 
\chi_{S_A M_{S_A}}^{123} \chi_{1-m}^{45} \right\rangle \notag \\
&\quad \times \left\langle L_A M_{L_A} S_A M_{S_A} 
\middle| J_A M_{J_A} \right\rangle  \left\langle L_B M_{L_B} S_B M_{S_B} 
\middle| J_B M_{J_B} \right\rangle \notag \\
&\quad \times \left\langle L_C M_{L_C} S_C M_{S_C} 
\middle| J_C M_{J_C} \right\rangle 
, \tag{B4}
\end{align}
where $E_i$ is the on-shell energy of hadron $i$, $\left\langle \varphi_C \varphi_B 
\middle| \varphi_A \varphi_0 \right\rangle$ denotes spin and flavor overlaps, and 
$I_{M_{L_B}, M_{L_C}}^{M_{L_A}, m}$ is the spatial integral given by
\begin{eqnarray}
I_{M_{L_B}, M_{L_C}}^{M_{L_A}, m}  &=& -3i\gamma \left(\frac{1}{2\pi}\right)^3 2^{-5/2} f^{-5}
\int \psi^*_{L_C M_{L_C}}(\mathbf{r}_C, \mathbf{R}_C) \notag \\
&& \times \psi^*_{L_B M_{L_B}}(\mathbf{r}_B) 
\mathcal{Y}_{1}^{m}(\mathbf{r}_\nu) 
e^{-i\mathbf{r}_{BC}\cdot \mathbf{p}} 
e^{-\frac{r_\nu^2}{4f^2}} e^{-\frac{R_{AV}^2}{R_0^2}} \notag \\
&& \times \psi_{L_A M_{L_A}}(\mathbf{r}_A, \mathbf{R}_A) 
d\mathbf{r}_A d\mathbf{R}_A d\mathbf{r}_\nu d\mathbf{R}_{AV}, 
\end{eqnarray}
where $\psi_{LM_L}$ are the spatial wave functions of the 
hadrons in coordinate representation, $\mathbf{r}_A$ and 
$\mathbf{r}_C$ are the relative coordinates between two 
quarks in baryons $A$ and $C$, $\mathbf{R}_A$ and 
$\mathbf{R}_C$ are the relative coordinates between the 
third quark and the remaining quark pair in baryons $A$ 
and $C$, $\mathbf{r}_B$ is the relative coordinate between 
the quark and antiquark in the meson, $\mathbf{r}_\nu$ is 
the relative coordinate of the created quark pair, 
$\mathbf{R}_{AV}$ is the relative coordinate between the 
initial baryon $A$ and the created quark pair, and 
$\mathbf{r}_{BC}$ is the relative coordinate between the 
final hadrons $B$ and $C$.

\bibliographystyle{elsarticle-num}
\bibliography{ref}

\end{document}